\documentclass[a4paper,11pt]{article}
\pdfoutput=1 

\usepackage{jinstpub} 

\title{NA62 Charged Particle Hodoscope. \newline Design and performance in 2016 run}

\author{S. Kholodenko}
\affiliation{Institute for High Energy Physics named by A. A. Logunov\\ of National Research Center "Kurchatov Institute",\\142281, 1 Nauki sq, Protvino, Russia}

\emailAdd{sergey.kholodenko@cern.ch}

\abstract{The NA62 experiment at CERN SPS is aimed to measure the branching ratio of the ultra-rare decay $K^+\rightarrow\pi^+ \nu \bar{\nu}$ with 10\% accuracy. The experiment operates with a 75~GeV/c high intensity~(750~MHz) secondary beam. A new detector, named Charged Particle Hodoscope~(CHOD), designed to produce an input signal to the L0 trigger processor for events with charged particles produced in kaon decays, has been assembled, installed, integrated in NA62 Data Acquisition System~(DAQ) and commissioned in 2016. During the whole 2016 run the detector has been in continuous operation. Design and performance features of the detector are presented.}


\collaboration[c]{on behalf of the NA62 collaboration$^{1}$\note{G.~Aglieri Rinella, R.~Aliberti, F.~Ambrosino, R.~Ammendola, B.~Angelucci, 
A.~Antonelli, G.~Anzivino, R.~Arcidiacono, I.~Azhinenko, S.~Balev
% ...
}}

\proceeding{Instrumentation for Colliding Beam Physics (INSTR-2017)\\
  2017\\
  Novosibirsk}

\begin{document}
\maketitle
\flushbottom

\section{Introduction}
\label{sec:intro}
\begin{figure}
\centering
\includegraphics[width=\linewidth]{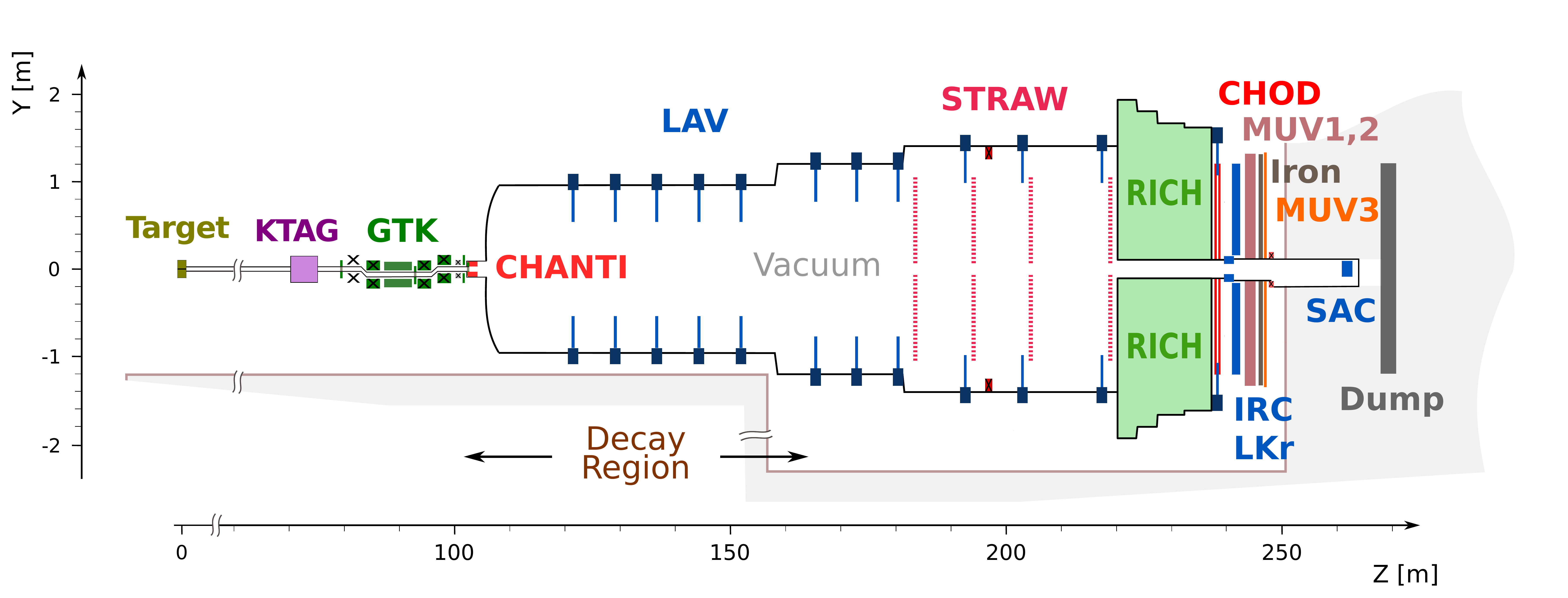}
\caption{Schematic view of the NA62 experimental layout}\label{fig:Setup}
\end{figure}

The NA62 experiment~\cite{DetectorPaper} is located at CERN North Area High Intensity Facility and is aimed to measure the branching ratio of the ultra-rare decay $K^+\rightarrow\pi^+ \nu \bar{\nu}$ with 10\% accuracy~\cite{Proposal}. The experiment operates with 75~GeV/c secondary beam comprising 6\% kaons at 750~MHz total rate. The schematic view of the experimental setup is shown on fig~\ref{fig:Setup}. The hardware Level-0~(L0) trigger processor, first step of the multi-level trigger system~\cite{NA62DAQ}, reduces the trigger rate down to 1~MHz. A new detector, named Charged Particle Hodoscope~(CHOD), provides an input to the L0 trigger based on the number of charged particles. The CHOD has been assembled and commissioned in 2016 and was in continuous operation during the whole 2016 data-taking period. 	

\section{NA62 CHOD detector}
\begin{figure}
\centering
\includegraphics[width=\linewidth]{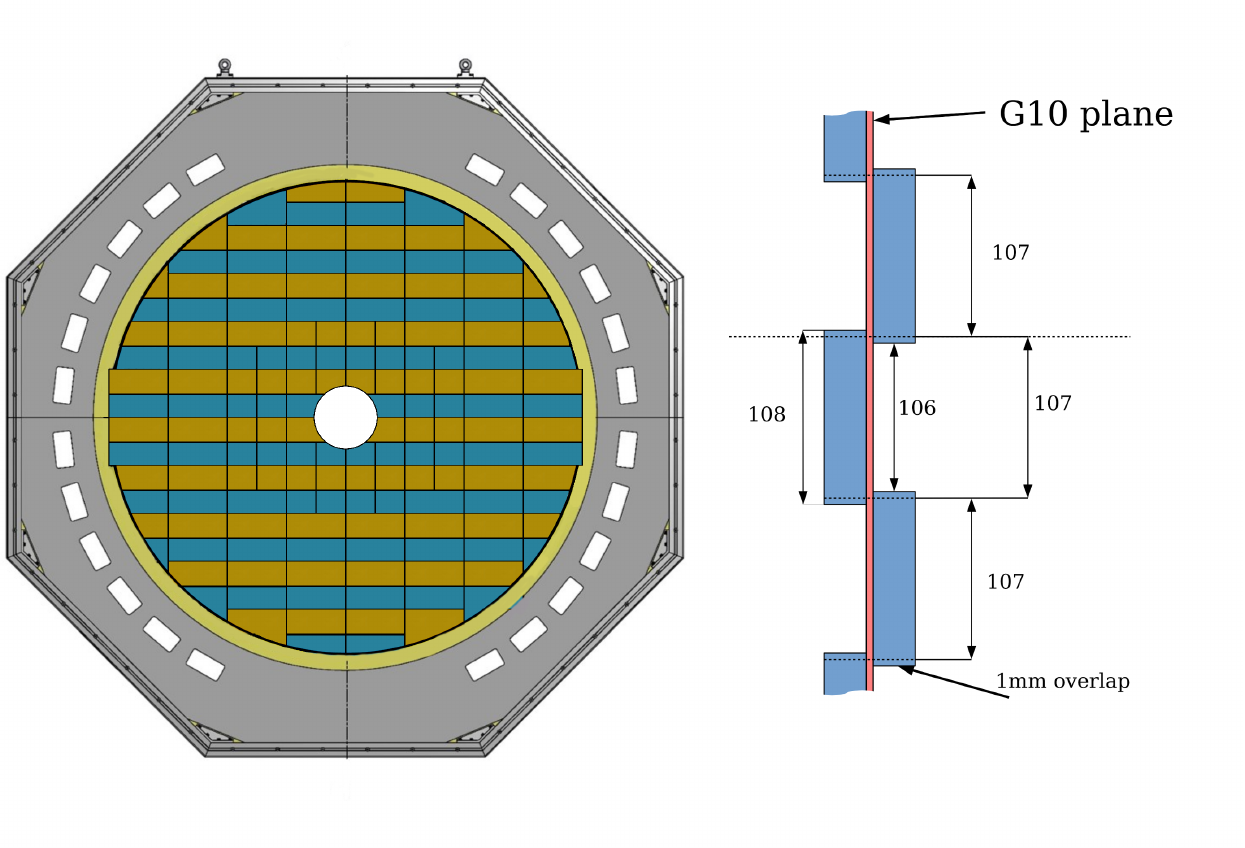}
\caption{NA62 CHOD design. Left: support panel with scintillator tiles. Right: schematic view of the tile rows layout}\label{fig:NewCHODSchematiView}
\end{figure}

The detector covers an annular area, 140~mm < R < 1070~mm, around the beam pipe~(see fig.~\ref{fig:NewCHODSchematiView}) and consists of scintillator tiles with Wave-Length Shifting~(WLS) fibre light collection and SiPM readout. The detector is located along the beam axis, after the RICH in front of the LAV12~(Large Angle Veto station 12) detector mainframe.

\vfil

\par 
The hodoscope consists of 152 scintillator, 30~mm thick and two typical sizes of $267.5 \times 108$~mm$^2$ and $133.75 \times 108$~mm$^2$ in the central region. Some peripheric tiles have a special shape to fit the circular acceptance area. Tiles are made of polymerized polystyrene scintillator SC-301 produced at IHEP~(Protvino, Russia)~\cite{ScintillatorProduction} and fixed on a 3~mm thick G10 support plane. The tiles are organized in rows: two contiguous rows are on opposite sides of the support plane with 1~mm~(fig.~\ref{fig:NewCHODSchematiView})  overlap. The efficiency to detect cosmic rays has been measured for each tile to be above 99\%, including edge effects~\cite{Semenov},\cite{MWPC}. 

\par Tiles are wrapped with a Tyvek sheet with the exception of adjacent faces which are interlaid with 70~$\mu$m thick double-Aluminized Mylar~\cite{Prototype}.

\par
\begin{figure}
 \begin{minipage}{0.45\textwidth}
  \centering
  \includegraphics[width=0.95\linewidth]{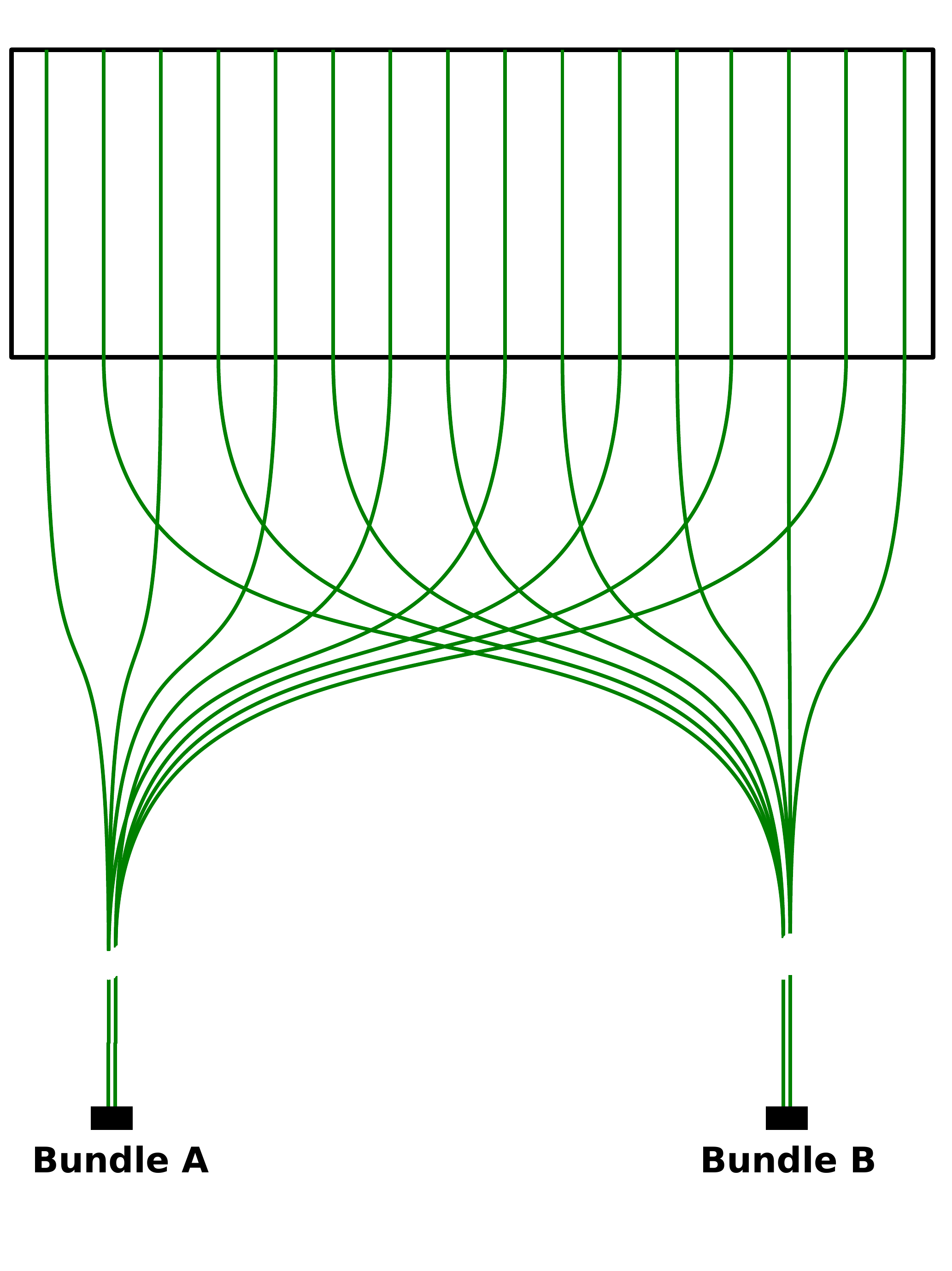}
  \caption{Schematic view of a rectangular tile $267.5 \times 108$~mm$^2$ readout via two bundles A and B of 8+8 interleaved WLS fibres.}
  \label{fig:TileReadout}
 \end{minipage}
 \hfill
 \begin{minipage}{0.45\textwidth}
   \centering
   \includegraphics[width=0.9\linewidth]{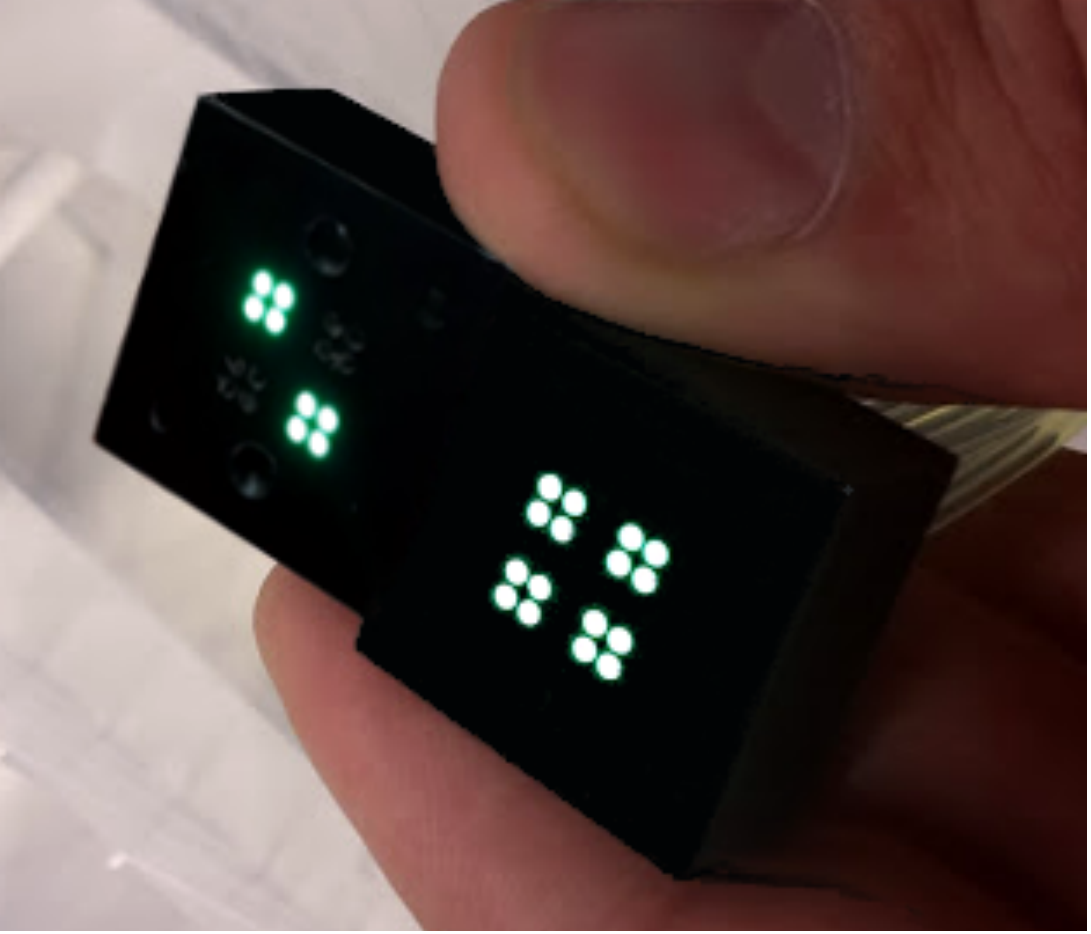}
   \caption{Photograph of the WLS-fibres holder. Fibres are organized in groups with 4 fibres each.}
   \label{fig:BlackCookie}
   \vfill
   \includegraphics[width=0.9\linewidth]{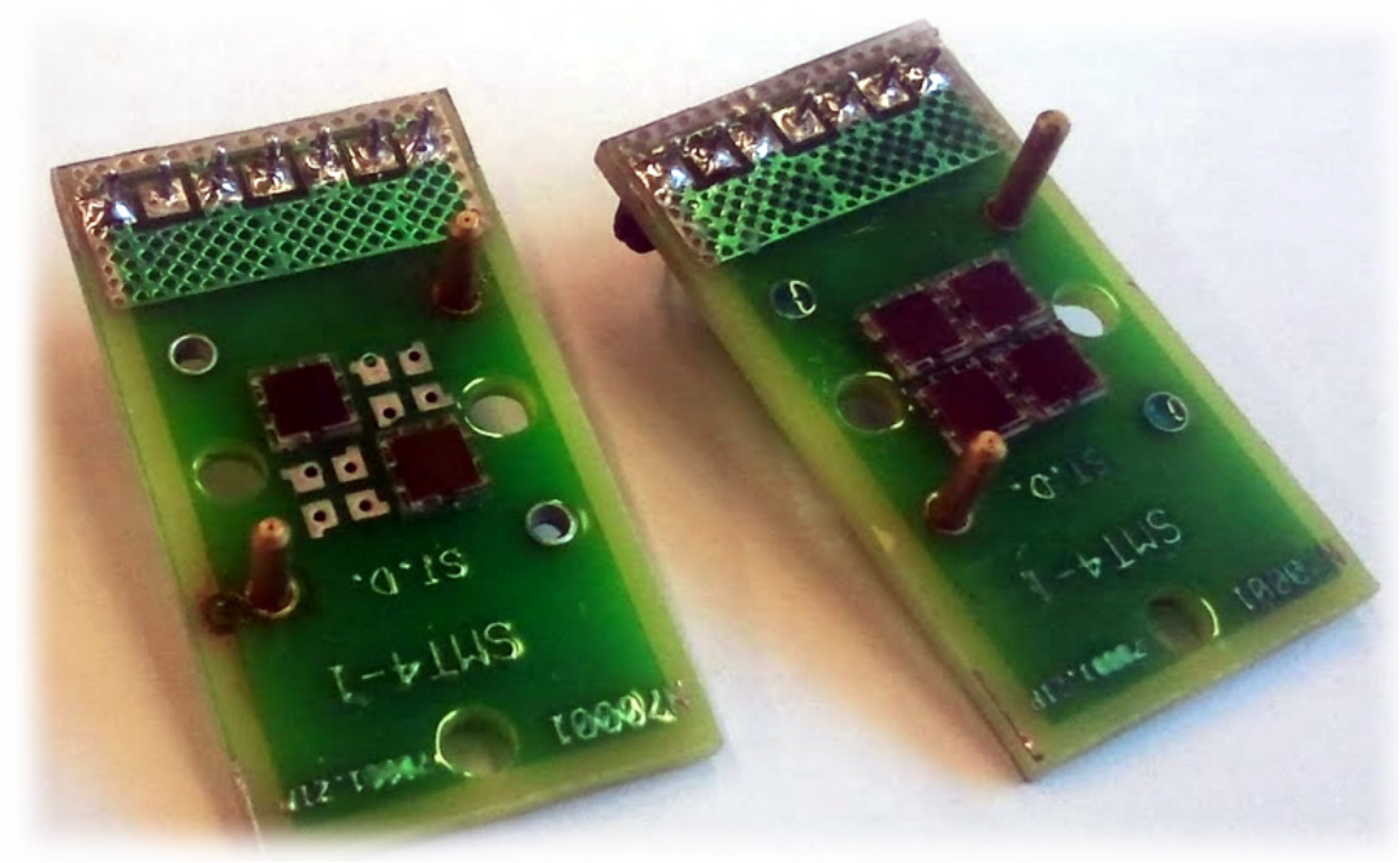}
   \caption{Photograph of the boards with SiPMs. Two SiPMs for small tiles and two groups of two SiPMs in series for large tiles.}
   \label{fig:SiPMPlate}
 \end{minipage}
\end{figure}

 The scintillating light from each tile is read by $\oslash 1$~mm wave-length shifting~(WLS) fibres Y11(200) type S that are organized in two bundles of interleaved fibres. Fibres are glued with EPO-TEK 301 Optical glue in grooves~(1.7~mm depth) parallel to the transverse 108~mm side of the tile. The distance between fibres is 16.71~mm or 33.42~mm between fibres from the same bundle~(the tile readout is schematically shown on fig.~\ref{fig:TileReadout}). Fibres were machined flat with a diamond milling tool at the two ends and sputtered with aluminum at the far end to increase the amount of light reaching the SiPM. To minimize the loss of light, four WLS-fibre lengths from 1350~mm to 2000~mm were chosen depending on the tile position.

\par To obtain good light collection uniformity the fibres in each bundle are organized into groups of 4 fibres per SiPM. Each group of 4 WLS-fibres are optically coupled to SensL MicroFC-30035~\cite{SensL} silicon photomultipliers~(SiPMs) with $3\times3$~mm$^2$ active area, which are installed on the outside periphery of the detector active area. To measure the total light transmitted by each bundle of fibres~(see fig.~\ref{fig:TileReadout}) two SiPMs are used with output in analog OR. The two ORs from a tile are separately preamplified and directed to discriminator~(LeCroy 4413).Their ECL outputs are translated to LVDS and read by TEL62 TDC boards~\cite{NA62DAQ}, which are used by the data acquisition system of the NA62 experiment.
\par During the 2016 data-taking period, the detector was operated with a bias voltage of $4.5~V$ above breakdown and the threshold were set at 9~excited pixels~(45~mV).

\begin{figure}
\centering
\includegraphics[height=200pt]{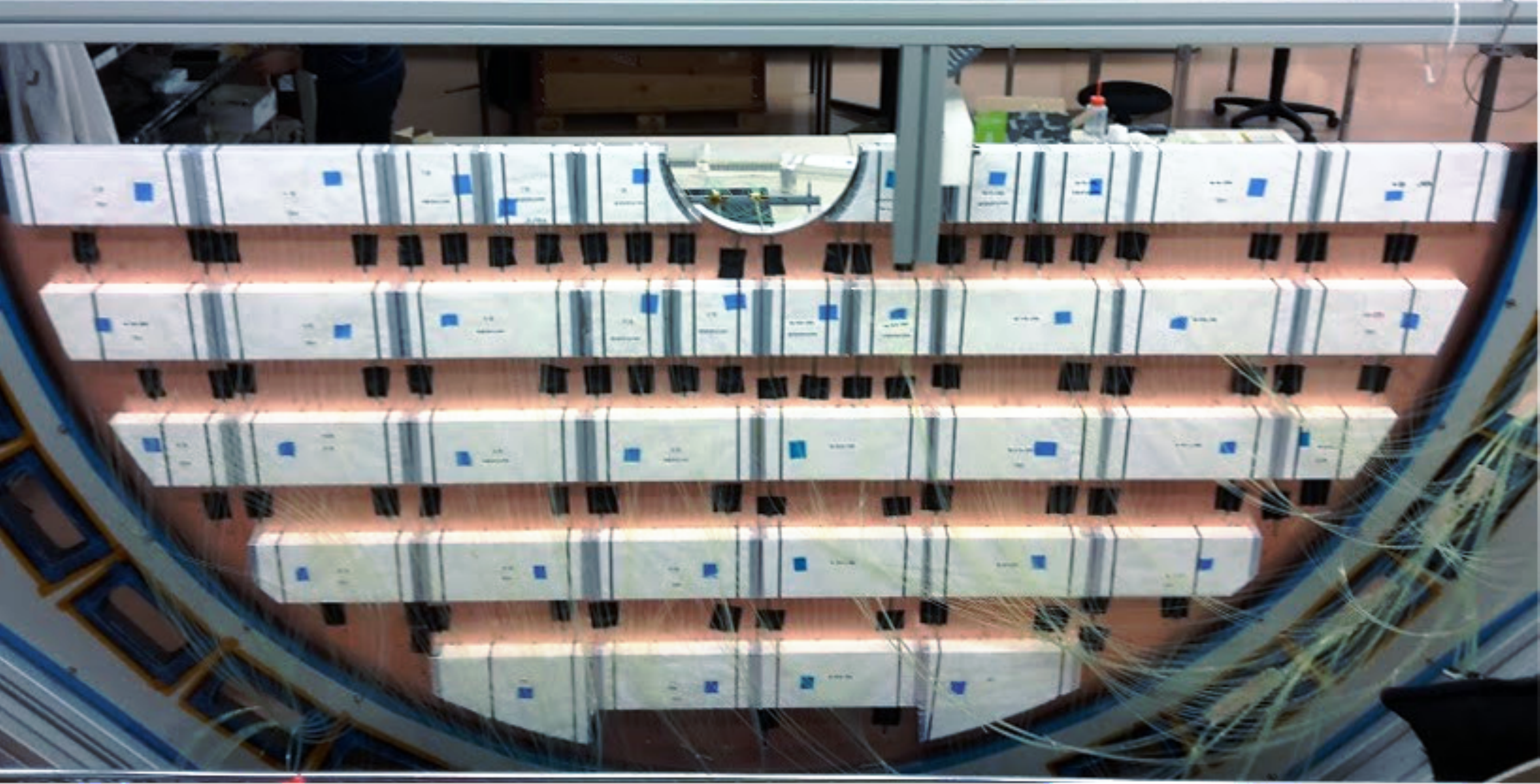}
\caption{Bottom half of the NA62 CHOD after assembly in the clean room and before transport to experimental hall}\label{fig:NewCHOD}
\end{figure}


\section{Performance during the 2016 data-taking period}
For efficiency and timing studies minimum bias topologies were used triggered by other detectors. Positive muons produced by $K_{\mu 2}$ decays~($K^+\rightarrow\mu^+ \nu$) were reconstructed from the spectrometer data and extrapolated to the CHOD plane.
\par
\begin{figure}
 \begin{minipage}{0.49\linewidth}
 \center {\includegraphics[width=\linewidth]{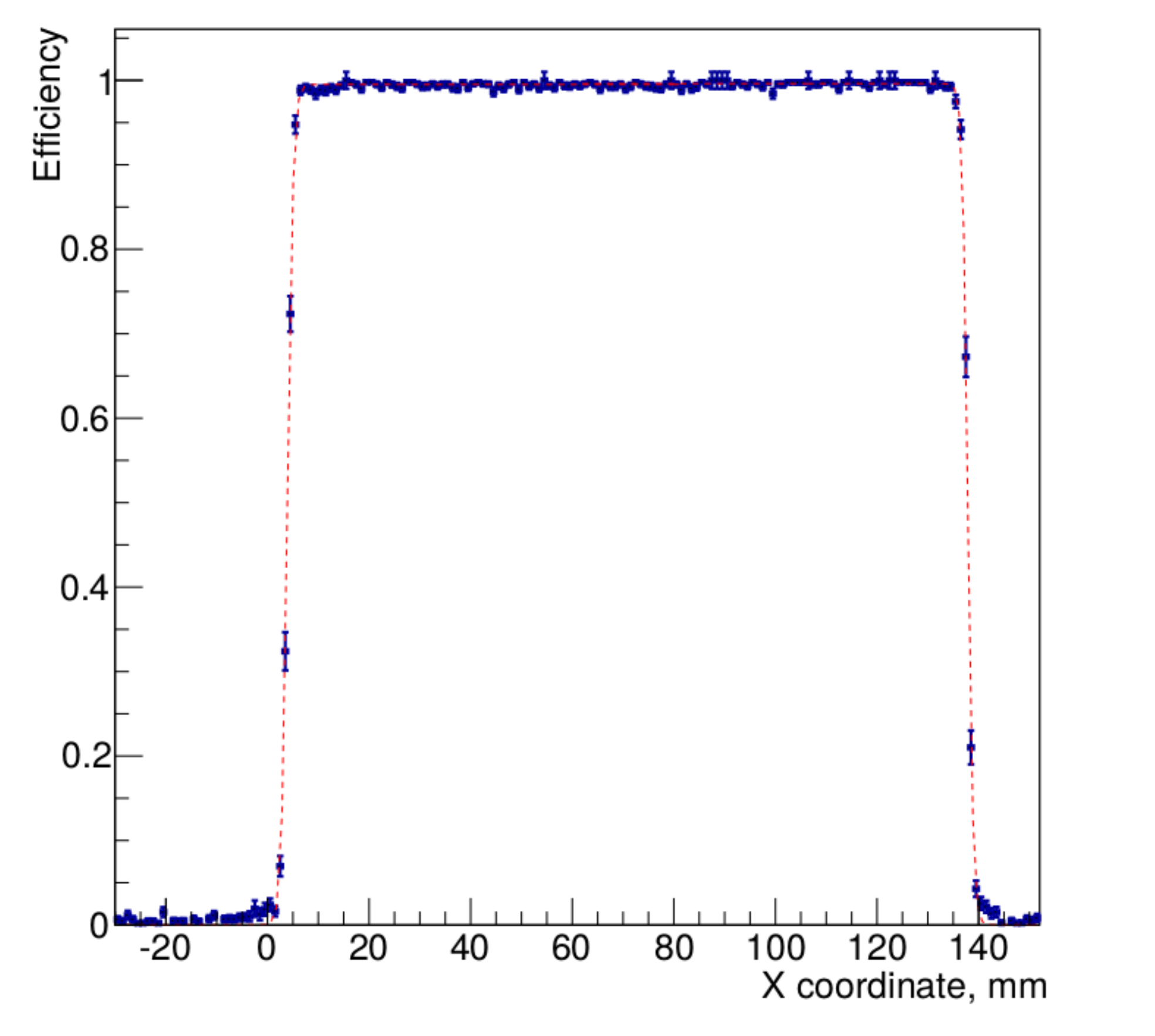}}
 \end{minipage}
 \begin{minipage}{0.49\linewidth}
 \center {\includegraphics[width=\linewidth]{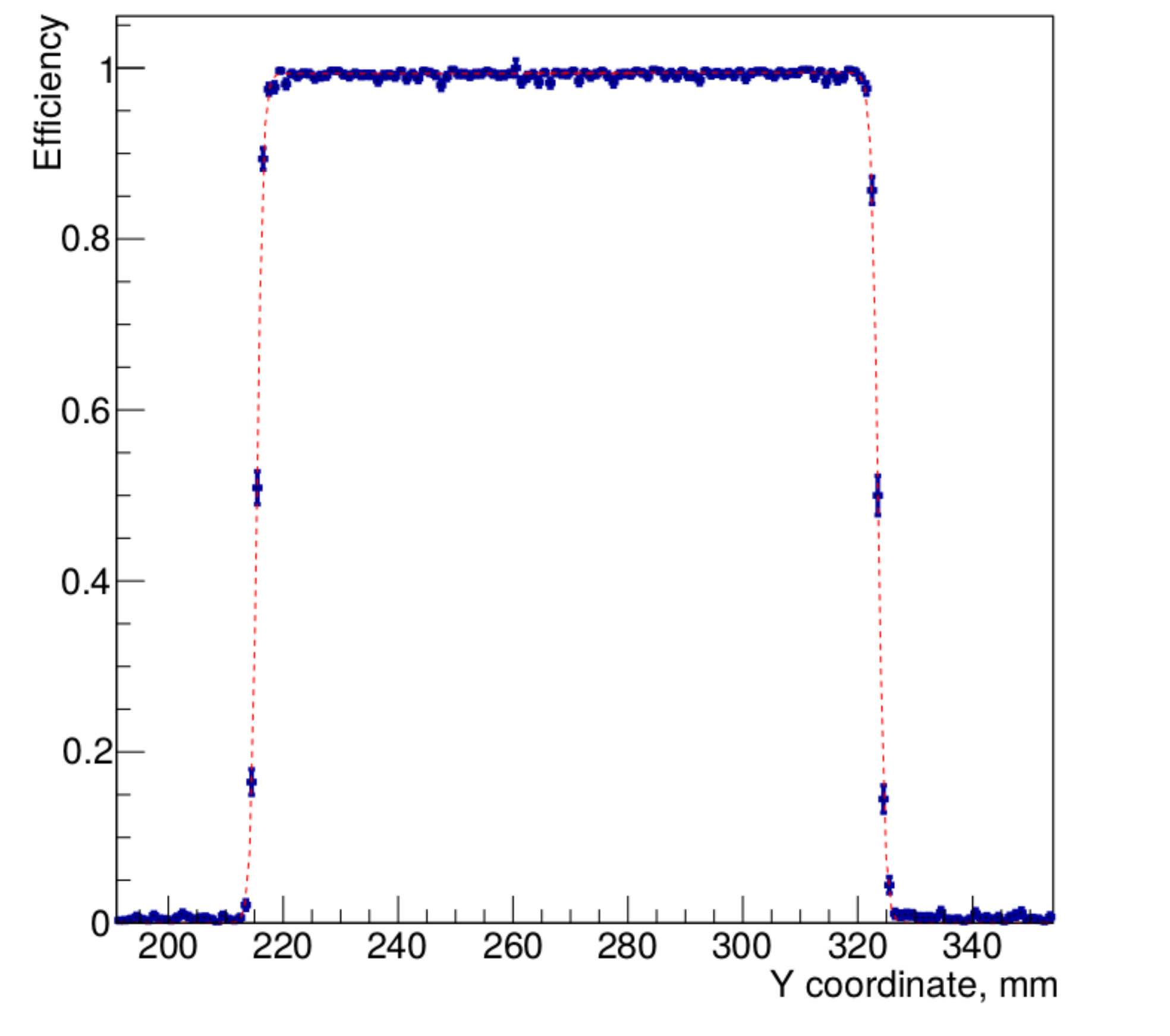}}
 \end{minipage}
\caption{Efficiency of a single tile as a function of the muon track extrapolated position on the CHOD surface: X coordinate~(left) and Y coordinate~(right). The dashed~(red) lines correspond to the fits using two Fermi functions.}
 \label{Fig:Tile112Efficiency}
\end{figure}

 The efficiency as a function of the transverse coordinates~(X,~Y) for a typical tile are presented on fig.~\ref{Fig:Tile112Efficiency}. The dashed red lines are fits with Fermi functions to the edges of the tiles. The plateau efficiency value obtained as a Fermi function parameter is $(99.5\pm0.1)$\%. Reconstructed tile sizes were estimated as $\Delta X = (133.96 \pm 0.05)$~mm and $\Delta Y = (108.11 \pm 0.05)$~mm assuming the edge of the tile as the point of 50\% efficiency. The physical size of this tile is $133.75 \times 108$~mm$^2$.
\par The overall average efficiency for the fiducial sensitive area of the detector was found to be $(98.6 \pm 0.1)$\%.

\par
\begin{figure}
 \begin{minipage}{0.48\linewidth}
 \centering
 \includegraphics[height=170pt]{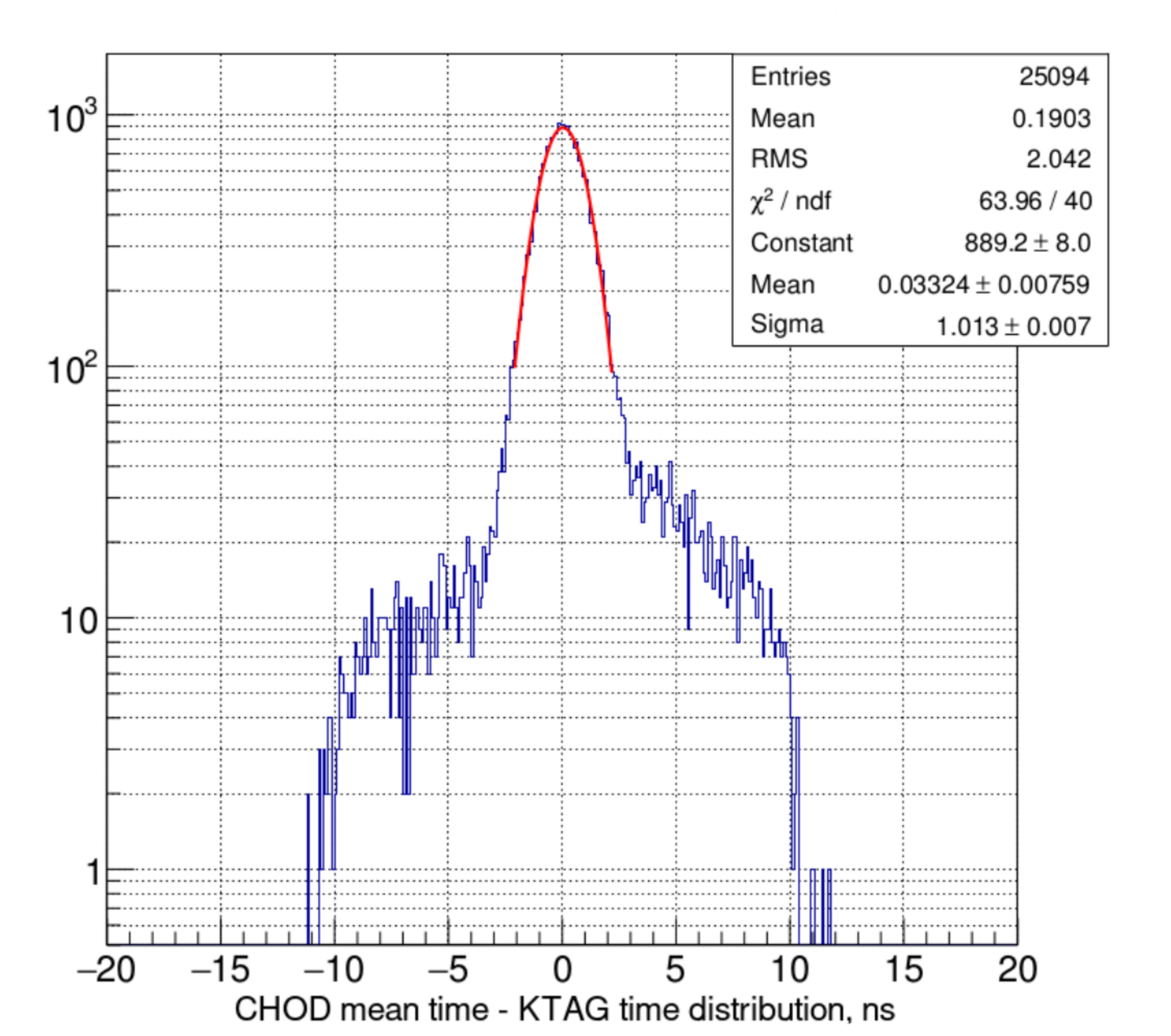}
 \caption{Mean signal arrival time distribution for the selected tile wrt KTAG. The red line corresponds to the Gaussian fit.}
 \label{fig:TimeResolution_single}
 \end{minipage}
 \hfill
 \begin{minipage}{0.49\linewidth}
 \centering
 \includegraphics[height=160pt]{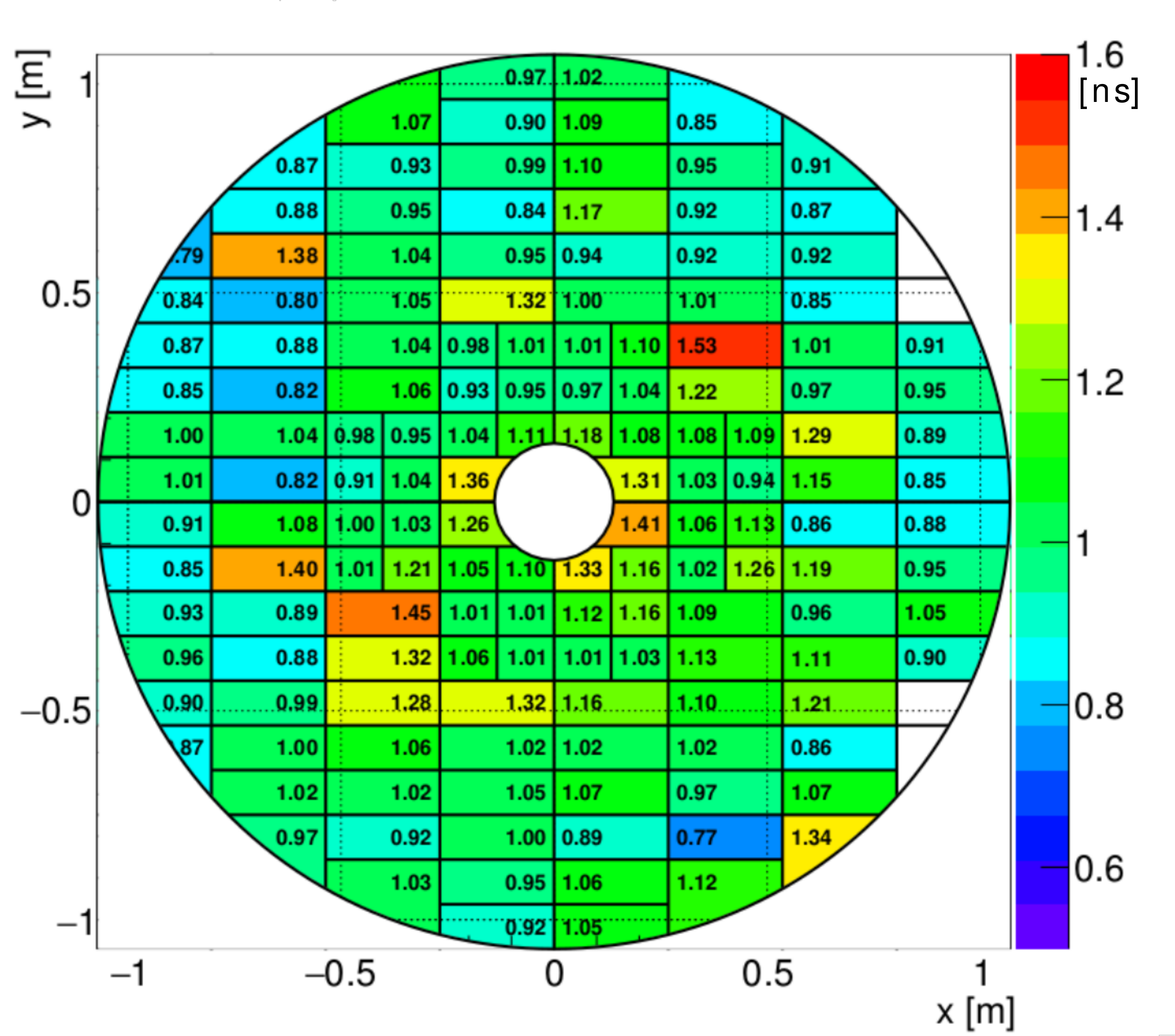}
 \caption{Time resolution~($\sigma$) of each of the NA62 CHOD tiles.}
 \label{fig:TimeResolution_All}
 \end{minipage}
\end{figure}

 For the time resolution measurements the differential Cherenkov counter~(KTAG), which selects kaons in the primary beam with a time resolution of $\sigma=80$~ps was used as a reference. Fig.~\ref{fig:TimeResolution_single} shows the average arrival time distribution of the A and B bundles of a tile with respect to the KTAG time. From the Gaussian fit, the time resolution~($\sigma$) is obtained to ($1.01 \pm 0.01$)~ns. The resolution of each tile is given on fig.~\ref{fig:TimeResolution_All} according to its position on the CHOD surface.

\section{Conclusion}
After installation and commissioning, during the whole NA62 2016 data-taking period, the CHOD operated reliably and was mainly used to provide the L0 trigger information on the multiplicity of tiles per event, with time resolution of ($1.1 \pm 0.1$)~ns and an efficiency averaged over $3.5$~m$^2$ sensitive area of $(98.6\pm 0.1)\%$. Both parameters are in agreement with the requirements of the experiment. Significant improvement of the time resolution is expected to be obtained by replacing the classical threshold discriminators by constant fraction discriminators~(CFDs) with the possibility of adjusting thresholds channel by channel. The prototype of the new CFD module has already been tested successfully.





\begin{thebibliography}{99}

\bibitem{DetectorPaper}
  E.~Cortina Gil [NA62 Collaboration],
  ``The Beam and detector of the NA62 experiment at CERN'',
  arXiv:1703.08501 [physics.ins-det]. Accepted for publication in JINST.

\bibitem{Proposal}
G. Anelli et al., 
  ``Proposal to Measure the Rare Decay $K^{+} \rightarrow \pi^{+} \nu \bar{\nu}$ at the CERN SPS'',
  CERN-SPSC-2005-013, SPSC-P326.
  \url{http://cds.cern.ch/record/832885/files/spsc-2005-013.pdf}

\bibitem{NA62DAQ}
B.~Angelucci {\it et al.},
  ``The FPGA based Trigger and Data Acquisition system for the CERN NA62 experiment'',
  JINST {\bf 9} (2014) no.01,  C01055.
  doi:10.1088/1748-0221/9/01/C01055

\bibitem{ScintillatorProduction}
Vladimir Rykalin, Valery Brekhovskikh, Sergey Chernichenko, Alexandre Gorin and Vitaliy Semenov, 
  ``Development of the Polystyrene Scintillator Technology and Particle Detectors on Their Base'', 
  Journal of Physical Science and Application5 (2015) 6-13,
  DOI:10.17265/2159-5348/2015.01.002

\bibitem{Semenov} 
V.~Semenov, V.~Brekhovskih, A.~Gorin, A.~Khudyakov, V.~Rykalin and O.~Yushchenko,
  ``Study of polystyrene scintillators-WLS fibre elements and scintillating tile-WLS prototypes for New CHOD detector of CERN NA-62 experiment'', PoS PhotoDet {\bf 2015} (2016) 041.

\bibitem{MWPC}
S.~A.~Kholodenko, A.~A.~Khudyakov, I.~Mannelli, V.~F.~Obraztsov, V.~D.~Samoylenko, V.~K.~Semenov and V.~P.~Sugonyaev,
  ``Time resolution measurements of scintillating counters for a new NA62 trigger charged hodoscope'',
  JINST {\bf 9} (2014) C09002.
  doi:10.1088/1748-0221/9/09/C09002

\bibitem{Prototype}
V.~Duk {\it et al.},
  ``Performance studies of the hodoscope prototype for the NA62 experiment'',
  JINST {\bf 11} (2016) no.06,  P06001.
  doi:10.1088/1748-0221/11/06/P06001

\bibitem{Y11}
Kuraray website: \url{http://kuraraypsf.jp/psf/ws.html}

\bibitem{SensL}
SensL website: \url{http://sensl.com/products/c-series/}






\end{thebibliography}
\end{document}